\documentclass[a4paper,11pt]{article}
\usepackage{amsmath,amsfonts,amssymb,mathtools}
\usepackage{url,xspace,a4wide}
\newcommand{\Andreka}{Andr{\'e}ka\xspace}

\newcommand{\Istvan}{Istv{\'a}n\xspace}

\newcommand{\Szekely}{Sz{\'e}kely\xspace}

\newcommand{\Tuple}[1]{\ensuremath{\left({#1}\right)}\xspace}
\newcommand{\Norm}[1]{\ensuremath{\mathopen{||}#1\mathclose{||}}\xspace}
\newcommand{\Mod}[1]{\ensuremath{\mathopen{|}#1\mathclose{|}}\xspace}
\newcommand{\wline}{\textsf{wline}}
\newcommand{\hop}[1]{\ensuremath{\xrightarrow{#1}}\xspace}
\newcommand{\jump}[2]{\ensuremath{{#1}\rightarrow{#2}}\xspace}
\newcommand{\Jump}[2]{\ensuremath{{#1}\Rightarrow{#2}}\xspace}
\newcommand{\st}{\mathrel{\mid}}
\newcommand{\ST}{\mathrel{.}}
\newcommand{\Implies}{\ensuremath{~~ \Rightarrow{} ~~}}

\newcommand{\QQ}{\ensuremath{\mathbb{Q}}\xspace}
\newcommand{\RR}{\ensuremath{\mathbb{R}}\xspace}

\newcommand{\HypRR}{\ensuremath{\mathopen{^*}\mathbb{R}}\xspace}

\newcommand{\amp}[2]{[\jump{#1}{#2}]}
\newcommand{\AMP}[2]{[\Jump{#1}{#2}]}

\newcommand{\SpecRel}{\ensuremath{\mathit{SpecRel}}\xspace}
\newcommand{\GenRel}{\ensuremath{\mathit{GenRel}}\xspace}
\newcommand{\AccRel}{\ensuremath{\mathit{AccRel}}\xspace}

\newcommand{\Ax}[1]{\textsf{Ax#1}\xspace}

\newtheorem{theorem}{Theorem}[subsection]
\newtheorem{lemma}[theorem]{Lemma}
\newtheorem{prop}[theorem]{Proposition}
\newtheorem{corr}[theorem]{Corollary}
\newtheorem{conj}[theorem]{Conjecture}
\newtheorem{defn}[theorem]{Definition}
\newtheorem{note}[theorem]{Note}
\newtheorem{example}[theorem]{Example}

\newcommand{\QED}{\hspace{\stretch{1}}$\Box$}
\newenvironment{proof}{\noindent \textbf{Proof.~}}{\QED\newline}

\newcommand{\Eq}[1]{(\ref{eqn:#1})}

\title{Modelling Quantum Theoretical Trajectories within\\Geometric Relativistic Theories}
\author{Mike Stannett\\
  Department of Computer Science, University of Sheffield\\
  Regent Court, 211 Portobello, Sheffield S1 4DP\\
  United Kingdom\\
  \url{m.stannett@dcs.shef.ac.uk}
}
\date{5 September 2009}

\begin{document}
\maketitle


\begin{abstract}
\Andreka and her colleagues have described various geometrically inspired first-order theories of special and general relativity, while \Szekely's PhD dissertation focuses on an intermediate logic of accelerated observers. Taken together, these theories provide an impressive foundation on which to build wider mathematical descriptions of physical reality, but they remain deficient in one important respect --- they do not include direct support for quantum theory. In this paper we will attempt to remedy this situation by incorporating a model of quantum theoretical trajectories that can reasonably claim to be physically meaningful.

We have recently shown that the `bidirectional model' of quantum trajectories --- in which particles are deemed to `hop' randomly from one spacetime location, $q$, to another, $q'$ (which can be either earlier or later in time than $q$), and in which paths comprise a finite number of hops --- is logically equivalent to Feynman's path-integral formulation when spacetime is assumed to be Euclidean. In this paper we extend the model to relativistic spacetimes, and argue that observers are subject to the same `quantum illusions' as in the Euclidean case --- for, even though motion is discrete and respects no built-in `arrow of time', observers have no choice but to perceive particle trajectories as continuous (locally) future-pointing paths in spacetime.

Whereas the relativistic theories presuppose continuous paths as part of their axioms, the `quantum illusion' of continuous motion allows us to replace this axiom with a lower-level quantum-inspired axiom concerning discrete jumps in spacetime. We investigate the nature of these jumps, and the extent to which they can be tied to the underlying geometric structure of spacetime. In particular, we consider hops of the form $q \to q'$ which preserve features of the underlying number field, and investigate the extent to which all hops can be restricted to be of this form.
\end{abstract}

\section{Introduction}
\label{sec:introduction}

What can we say about the way bodies move in space and time? We'll begin by considering this question in the context of inertial bodies subject to special relativistic kinematics as formalised in \cite{AMN04,AMN07,Sze09} by the theory \SpecRel over $(1+N)$-dimensional spacetime, where $N > 1$ (e.g.~taking $N=3$ corresponds one temporal dimension and three spatial ones).

\subsection{The axioms of \SpecRel}
\label{sec:The-axioms-of-SpecRel}

What do we actually perceive when we observe an inertial object moving from one location to another? The usual answer is implicit in the axiom
\begin{quote}
\Ax{Line}: $(\forall m \in Obs, \forall h \in Obs \cup Ph)$ \quad $tr_m(h) \in Lines$
\end{quote}
which says that each (inertial) observer $m$ considers the trajectory of every (inertial) observer or photon $h$ to trace out some straight line. In particular, $m$ sees herself stationary in space, as indicated by the axiom
\begin{quote}
\Ax{Self}: $(\forall m \in Obs)$ \quad $tr_m(m) = \bar{t}$
\end{quote}
which says that $m$ considers her worldline to be the time axis $\bar{t} \equiv \{\Tuple{q, 0, \dots, 0} \st q \in Q \}$ in $Q^{N+1}$. The entity $Q$ is the (algebraic) field of values used to coordinatize spacetime, which in \SpecRel and its extensions (\AccRel and \GenRel) is assumed to satisfy
\begin{quote}
\Ax{Field}: $Q$ is a linearly ordered Euclidean field, i.e. every positive $q \in Q$ has a square root in $Q$.
\end{quote}

What about the other axioms of \SpecRel? Recall that
\[
	\SpecRel \equiv \{ \Ax{Line}, \Ax{Self}, \Ax{Pot}, \Ax{Events}, \Ax{Ph}, \Ax{Sym} \}
\]
where, in the notation of \cite{AMN04}\footnote{
	The notation of \cite{AMN07} is slightly different, and the axioms are expressed 
	slightly differently. In particular, traces $tr_m(k)$ are renamed worldlines $\wline_m(k)$, and the 
	axioms \Ax{Pot} and \Ax{Ph} are refactored: \Ax{ThEx} replaces \Ax{Pot} for 
	observers, and \Ax{Ph} is extended to include the other half of \Ax{Pot}.
}, and for all $m,k \in Obs$, $ph \in Ph$ and $p,q \in \bar{t}$,
\begin{quote}\begin{tabular}{ll}
	\Ax{Pot}:    & $ \left( ang(l) < 1 \Rightarrow (\exists k \in Obs)(l = tr_m(k)) \right) \land$ \\ 
	             & $ \left( ang(l) = 1 \Rightarrow (\exists ph \in Ph) (l = tr_m(ph)) \right)$ \\
	\Ax{Events}: & $Rng(w_m) = Rng(w_k)$ \\
	\Ax{Ph}:     & $v_m(ph) = 1$ \\
	\Ax{Sym}:    & $ \Norm{f_{mk}(p)_t - f_{mk}(q)_t} = 
	                 \Norm{f_{km}(p)_t - f_{km}(q)_t}$
\end{tabular}\end{quote}
These capture, respectively, the following notions:
\begin{itemize}
\item
	\Ax{Pot} says that if a straight line, $l$, lies within (or on) a lightcone, then there exists a [potential] observer (or photon, respectively) $k$ for which that line coincides with $tr_m(k)$;
\item
	\Ax{Events} says that all observers see the same events (though possibly at different coordinates);
\item
	\Ax{Ph} says that all observers consider the lightcone angle to be 45$^\circ$ (i.e. photons always travel with speed $1$);
\item
	\Ax{Sym} says that mutual observers see each other's clocks running at the same (necessarily \emph{slow}) rate.
\end{itemize}

Notice, however, that none of these axioms requires motion to be continuous from a higher-dimensional point of view, because the theory is inherently \emph{geometric}; bodies are represented not by moving points but by \emph{completed traces}. It matters that every point of the trajectory $tr_m(h)$ is seen by $m$ to be visited by $h$, but the order in which the points along $tr_m(h)$ are populated is not constrained in any way by any of the axioms of \SpecRel.

We shall exploit this freedom, using it to redefine the nature of higher-dimensional motion in a way that remains consistent with \SpecRel, but which at the same time allows us to model paths as superpositions of \emph{finite computations}. This will in turn allow us to clarify the meaning of certain constructions in \SpecRel and its extensions\footnote{
	The approach adopted here is essentially just an extension of our paper \cite{Sta09}, in which finitary motion is shown to generate a model that is formally equivalent in a Euclidean setting to Feynman's path-integral formulation of quantum mechanics \cite{Fey48}.
}.

\subsection{Discrete finitary motion}
\label{sec:discrete-finitary-motion}

The type of motion we'll be describing in this paper is \emph{discrete} and \emph{finitary}, by which we mean that bodies move in a series of discrete spacetime \emph{hops}, where a hop takes the body from one location, $q = (t,\bar{x})$, to another, $q' = (t', \bar{x}')$, without passing through any intermediate points: we indicate such a hop by writing $q  \to q'$, or if we want to assign the hop a label, $h$, we write $q \hop{h} q'$. All of the hops described in this paper treat space and time symmetrically, in the sense that the particle can hop into either its future \emph{or its past} (we do not assume that $t'$ is later than $t$), in the same way that it can move in any spatial direction. A \emph{finitary path} (or just \emph{path}) from $q$ to location $q'$ is then a \emph{finite} string of hops, $ q = q_0 \hop{h_1} \dots \hop{h_n} q_n = q $, and we will sometimes write $q \hop{p} q'$, where $p = h_0 \dots h_n$. There are obviously many paths from one location to another, and we assign each one a complex number (its \emph{amplitude}), as explained in \ref{sec:amplitudes-and-the-hop-action} below.\footnote{
	We follow Feynman in calling these complex numbers \emph{probability 
	amplitudes}, or just \emph{amplitudes} for short; if $p$ has amplitude $z = r e^{\imath\theta}$, 
	then $r^2$ indicates the probability that the path $p$ will be traversed. 
	Readers should beware that some authors use the word `amplitude' to 
	refer to the magnitude $r$ rather than the value $z$; we shall never do so.
} This will then enable us to carry out simple quantum theoretical calculations.

\subsubsection{Motivation}
\label{sec:motivation}

Before giving the details, we ought first to motivate these definitions. Our position is essentially solipsist: perception is an active process by which we project structure onto otherwise unstructured sense-impressions. It follows from this position that we cannot know how bodies `actually' move --- all we can say is how we \emph{perceive} them to move. However, we should not forget that what we perceive depends on where we position ourselves. The observers modelled within a \SpecRel universe need not agree with us when it comes to describing the motion of inertial bodies, because we stand \emph{outside} the model in order to reason about it; in effect, we are not observers, but `super-observers'.

As we have seen, \Ax{Line} tells us that the worldline of an inertial body (or a photon) is a straight line in $Q^{N+1}$. Since it is a theorem of \SpecRel that no photon or inertial observer can travel faster than light\footnote{
	It is shown in \cite[Theorem~2]{AMN04} that the theory $\SpecRel_0$ over 
	$d$-dimensional spacetime entails no faster-than-light observers, 
	provided $d>2$. Since $\SpecRel \equiv \SpecRel_0 \cup \{\Ax{Sym}\}$ and 
	$d = (1+N) > 2$, the same holds true in the models of \SpecRel considered here.
	}, the worldline $tr_m(k)$ must lie either within or on a lightcone. In particular, therefore, $tr_m(k)$ specifies a continuous injective function of type $Q \to Q^N$, mapping each point of $m$'s time axis to a corresponding set of spatial coordinates, and it is natural for $m$ to consider $k$ to be moving along a continuous spatial path as time passes.
	
Suppose, however, that $m$ `actually' travels along her own worldline (by \Ax{Self} this is the time axis) in the following disjointed fashion: she starts by moving smoothly forwards in time until $t =4$, then she reverses direction and moves smoothly backwards through time to $t=3$, then reverses direction a second time and moves forward through time thereafter. From our perspective as super-observers, we know that $m$ traversed the time interval $[3,4]$ three times, twice moving forward through time, and once backwards. But $m$ cannot possibly detect this herself, because the events --- the only things on which $m$ can base her worldview --- occurring at each repeated instant (e.g.~$t = 3.5$) are the same each time she arrives there, because \Ax{Line} requires the three sub-paths to coincide. Standing outside the model, we can consider the order in which different parts of $tr_m(m)$ become populated; $m$ has no such luxury, since (by definition) she has only one way to model time, namely as her position along the axis $\bar{t}$. As this example shows, we can use our `super-observer rights' to consider a wide range of distinct dynamics, all of which appear to be identical from $m$'s point of view. 

However, we take this argument further: we assert that $m$ cannot possibly perceive any trace $tr_m(k)$ in its entirety, because this would imply the (arguably unreasonable\footnote{
	This issue needs further investigation. We are thinking of $Q$ as a subfield 
	of $\RR$, so that all non-empty subintervals have finite length. However, the axioms 
	of \SpecRel also allow models of $Q$ based on the hyperreal line \HypRR, and it is 
	not immediately clear why observation times could not be infinitesimal in that setting.
	}) ability to perform infinitely many distinct observations in finite time (from $m$'s point of view). How else, for example, could $m$ perceive every single instant of the time interval $[0,1] \in \bar{t}$?

Assuming that only finitely many observations can be made during any finite interval, it follows that when $m$ observes $k$'s motion, all $m$ actually sees is a finite subset of $tr_m(k)$. But as we have seen, we have the freedom, as super-observers, to question the order in which those finitely many observed positions are actually instantiated --- and, of course, our choices in this matter \emph{should have no effect on the way $m$ perceives things}. Equally, the fact that $m$ makes the observations one after another (i.e. relative to her interpretation of temporal order) isn't relevant from our point of view, and we have the freedom to consider these observations to be made in any order, and to be repeated as often as we like. It is precisely this representation that is expressed using our finitary path concept --- bodies jump from one location to another without passing through intervening points, and this motion can take them both backwards and forwards along $m$'s time axis.

Our goal will be to show that each trace can be approximated arbitrarily closely by a converging sequence of (essentially random) finitary paths, each with an associated amplitude. We replace the standard notion of continuous motion by a quantum superposition of these finite approximations, and show that the original traces of \SpecRel are necessarily recovered in their entirety. This will enable us to apply standard principles of (algebraic) field theory to worldlines, provided we restrict attention to finitary paths whose coordinates, $(q, \dots, q')$, can be given as the roots of some suitable common polynomial. We explain this in more detail below, and show that the restrictions in question can indeed be applied in a way that remains consistent with the axioms of \SpecRel.

\subsubsection{Amplitudes and the hop action}
\label{sec:amplitudes-and-the-hop-action}

The procedure we follow is essentially identical to that used by Feynman to derive the standard path-integral formulation, except that the paths we consider are somewhat different from those of Feynman's formulation. The whole point of our construction is to \emph{deduce} that the path observed when $m$ watches $k$ is the trace $tr_m(k)$; we cannot, therefore, simply assume that $k$ moves along this path. Instead, we allow the path followed by $k$ to be essentially \emph{random}, and then use properties of quantum superposition to show that when these random paths are summed, the resultant motion nonetheless appears, from $m$'s point of view, to coincide with $tr_m(k)$. 

We begin by assuming that each hop $q \hop{h} q'$ has an \emph{action} $s(h) \equiv s(\jump{q}{q'})$, a real value, associated with it. The amplitude $\amp{q}{q'}$ associated with $h$ is then the complex number
\begin{equation}
	\amp{q}{q'} = B_0 e^{-is(\jump{q}{q'})/\hbar} \label{eqn:hop_amp}
\end{equation}
where $B_0$ is a normalisation factor, and $\hbar$ is the reduced Planck constant. The physical meaning of choosing $B_0$ to be a global constant is that all hops are equally likely to occur. To compute the amplitude of a path $p = h_1\dots h_n$, we define
\[
	[h_0 \dots h_n] = B_{n} e^{ -i(\sum_{j=0}^{n}{s(h_j)})/\hbar }
\]
where $B_n$ is again a normalisation factor; we regard all paths of length $n$ as being equally likely to occur at random, but they need not have the same chance of occurring as paths of some other length. The amplitude $\AMP{q}{q'}_n$ that a particle travels from one location $q$ to another location $q'$, via a path of length $n$ (lying entirely within a given region $R$) is then given by integrating over all possible paths. Finally, the amplitude $\AMP{q}{q'}$ that the particle does the journey via any permissible finitary path is given by summing over path lengths
\[
	\AMP{q}{q'} = \sum{ \AMP{q}{q'}_n }
\]

\subsection{Extension to \SpecRel}
\label{sec:extension-to-SpecRel}
In the same way that we have introduced a normalisation factor $B_n$ for hop-based paths of length $n$, so Feynman introduces a normalisation factor $A_n$ for continuous paths passing through $n$ points equally spaced in time. We have shown in \cite{Sta09} that the two interpretations will give identical results for all amplitude calculations (and hence that they are mathematically and observationally equivalent) provided we define
\begin{eqnarray}
	s(\jump{q}{q'}) &=& S(\jump{q}{q'}) + \rho \hbar \label{eqn:action} \\
		              & & \text{ where $q'$ occurs later than $q$ and $e^{i\rho}  = A_0 / B_0$ } \nonumber
\end{eqnarray}
where $S$ is the \emph{classical action}. Recall that $S$ is defined classically in terms of the \emph{Lagrangian} $\mathcal{L}$ of the system (the details need not concern us) by
\[
	S = \int_P{ \mathcal{L} } ~ dt
\]
where $P$ is the path followed by the object during the time in question. It is an extremely important physical principle that the (necessarily continuous) path followed by a classical particle is one which minimises $S$.

\begin{corr}
\label{corr:classical-action-continuous}
The classical action $S$, and hence the hop action $s$, are continuous functions of their end-points. \QED
\end{corr}

The value of $\rho$ is undetermined in our model, and has no observational significance; it represents the action of the \emph{null hop} (any hop of the form \jump{q}{q} whose source and target are identical; Feynman's model doesn't include subpaths of this kind). The requirement that $q'$ occur later than $q$ in \Eq{action} is a technicality, but presents no difficulties. Given a past-directed hop \jump{q}{q'}, we regard it instead as a \emph{forward}-directed hop made by the particle's anti-particle, and perform the calculation accordingly.

The amplitudes specified in \cite{Sta09} can be expressed recursively,
\[
	\AMP{q}{q'}_n
			= \frac{B_{n-1}}{B_n} \int_{X_n}{ \int_{T_n}{ \AMP{q}{\Tuple{t,\bar{x}}}_{n-1} \amp{\Tuple{t,\bar{x}}}{q'} \; dt \; d\bar{x} }}
\]
where $X_n$ and $T_n$ specify the ranges over which each hop's spatial and temporal coordinates are assumed to range. To keep things simple, we usually assume that $R$ is a basic (hyper)rectangle, $R = X \times T$, where $T$ is an interval $T = [T_{min}, T_{max}]$. If we wish to restrict attention to `unidirectional' non-relativistic processes (ones which involve only future-pointing hops), we would take $X_n = X$ and $T_n = [t_n, t_{max}]$ (since any future-pointing hop is permissible). In the `bidirectional' model (in which bodies hop in both temporal directions) we allow hops to any point in time, so we take $T_n = T$.

Clearly, we can adapt this recursive solution to any model of spacetime in which it is is possible to specify which regions of spacetime are accessible from each current location $q$. Indeed, we have two mechanisms for doing so. For \SpecRel and its extensions, for example, we can either restrict $R$ to include only the interior of the the lightcone\footnote{
	Claiming that hops only occur within lightcones is similar to asserting 
	that faster-than-light travel is impossible. However, we have to be careful
	not to draw an exact analogy, for while we can certainly consider the `slope'
	of an instantaneous hop, it is unclear in what sense this can be described 
	as its `speed'.
} at $q$, or else check (as must be the case if classical relativity is to be compatibility with these theories) that the classical action $S$ already ensures that hops across a lightcone boundary occur with vanishingly small probability.

\section{Restriction to Dense Subfields and Algebraic Hops}
\label{sec:algebraic-hops}

As with all theories, the usefulness of \SpecRel relies on the relevance of its axioms, and the most fundamental of these is \Ax{Field}, the assertion that $Q$ is an ordered Euclidean field. However, $Q$ is primarily used to coordinatize spacetime, whereas our focus is on \emph{observations}. This presents no problems to theories like \SpecRel, provided we assume that observed non-geometric values, like mass, also range over $Q$, but in `real life' not even the most accurate physical measurements can be made to more than a few decimal digits: when physicists measure masses, lengths and durations, the results typically range over the \emph{non-Euclidean} field \QQ. Two questions naturally arise in this context:
\begin{itemize}
\item
	what happens if we adopt an \emph{experimentalist} viewpoint, and assert that the observed universe is `really' modelled not by $Q^{N+1}$, but by $D^{n+1}$ for some possibly smaller field $D$? Since $D$ is used to provide measurements of comparable quantities like mass and distance, we will still require $D$ to be ordered. But we want more than this, because we traditionally think of physical measurements (perhaps mistakenly) as \emph{approximations} to some `true' underlying value, made `to within experimental error'. In other words, we want $D$ to be \emph{topologically dense} in $Q$, i.e. every non-empty open interval in $Q$ should contain an element of $D$.
\item
	is there any place left for $Q$ if we model everything using $D$? Our everyday experience tells us that $Q$ represents some kind of `ideal' coordinate space, while observations in $Q$ are only approximations; what we need, therefore, is some natural way of recapturing the idealised $Q$-based worldview from the lower-level $D$-based reality. This is, in part, the role played by the hop-based discrete finitary motion described in this paper.
\item
	although \Ax{Field} allows $Q$ to be non-Archimedean (for example, taking $Q$ to be the hyperreal field \HypRR is compatible with \Ax{Field}), is it meaningful to extend this freedom to $D$? Everyday experience suggests not. For example, we would expect to generate any required mass simply by piling together a large enough (but still finite) pile of 1kg shop-weights. We leave this question open.
\end{itemize}

\subsection{Square-dense fields}
\label{sec:square-dense-fields}

Since we are now thinking of $D$ as the field of `observable values', contained in the Euclidean field $Q$ of `ideal values', we need to ask: given an ordered field $D$, when can we find an ordered Euclidean extension $Q$ in which $D$ is topologically dense?

Suppose $E/F$ is a field extension (i.e. $F \leq E$). We'll say that $F$ is \emph{square-dense in} $E$ (or that $E/F$ is \emph{square-dense}) provided every $E$-interval that contains a positive element of $F$ contains an $F$-square, i.e.
\begin{quote}
  $\Ax{Root$_2(E/F)$}$: $\forall a,b \in E \ST
     \left(
     	[\exists c \in F \ST (0 < a < c < b)] \Implies
     	[\exists d \in F \ST (a < d^2 < b)]
     \right)$
\end{quote}
In other words, each positive element of $F$ can be approximated arbitrarily closely by squares (from $F$); but the definition of `arbitrarily closely' depends on the extension field $E$ being considered. If $F$ is square-dense in its real-closure,\footnote{
	Given any ordered field $F$, the Artin-Schreier theorem says that $F$ 
	has an essentially unique real-closed algebraic extension $R$. The field 
	$R$ is called the \emph{real-closure} of $F$.
} we'll simply say that $F$ is \emph{square-dense}. Notice that square-density is `inherited' by sub-extensions:

\begin{prop}
\label{prop:sd-subextension}
	Suppose $D \leq K \leq L$ are fields, and that $D$ is square-dense in $L$. Then $D$ is also square-dense in $K$.
\end{prop}
\begin{proof}
	Let $0 < k_1 < k_2$ in $K$ and suppose $d \in (k_1, k_2)_K \cap D$. Since the order topology on $K$ is inherited from that on $L$, we know that $(k_1, k_2)_K = (l_1,l_2)_L \cap K$ for some $l_1, l_2 \in L$. Since $D$ is square-dense in $L$ and the interval $(l_1,l_2)$ contains $d$, it must also contain $e^2$ for some $e \in D$. But $e^2 \in D \leq K$, so $e^2 \in (k_1,k_2)$ as required.
\end{proof}

\noindent Having defined square-dense fields, we show they exist and consider their properties.

\begin{example}
\label{ex:Q-is-sd}
\normalfont{\QQ is square-dense.}
\end{example}
\begin{proof}
Note first that \QQ is square-dense in \RR. This is a consequence of the Babylonian method for computing square roots. Given any positive rational $q$ we construct a rational sequence \Tuple{x_n} converging to $\sqrt{q}$ by setting $x_0 = 1$ and $x_{n+1} = \frac{1}{2}\left(x_n + \frac{q}{x_n}\right)$. But now $x_n^2 \to q$, whence any interval containing $q$ also contains some $x_n^2$. But the real closure of \QQ is the field of algebraic real numbers, and this is a sub-extension of \RR. It follows by Proposition \ref{prop:sd-subextension} that \QQ is square-dense.
\end{proof}

\begin{example} \normalfont{
	Every ordered Euclidean field is square-dense (because every positive value is itself a square), but the converse is not true; for example, \QQ is square-dense (Example \ref{ex:Q-is-sd}) but not Euclidean ($\sqrt{2}$ is irrational).} \QED
\end{example}

\begin{example} \normalfont{
	It is possible for a field to be square-dense in itself, but not square-dense in all of its extensions. For example, the hyperreal field \HypRR is ordered, so must have characteristic zero. Consequently it extends the square-dense field \QQ. Choose an infinitesimal value $\epsilon \in \HypRR$, and consider the interval $(2-\epsilon,2+\epsilon)$ in \HypRR. This interval contains only one rational, namely the value $2$, but this is not a square in \QQ. } \QED
\end{example}

\begin{lemma}
\label{lemma:dense-and-sd}
Suppose $K$ is dense in $L$. Then $K$ is square-dense in $L$ if and only if $K$ is square-dense in itself.
\end{lemma}
\begin{proof}
To see that $K$ square-dense in $L$ implies $K$ square-dense in itself, consider the extensions $K \leq K \leq L$ and apply Proposition \ref{prop:sd-subextension}. Conversely, suppose $K$ is square-dense in itself, and choose $0 < a < b$ in $L$. Since $K$ is dense in $L$, there exists $d \in K \cap (a,b)$, and again by density there exist $a', b' \in K$ such that $a' \in (a,d)$ and $b' \in (d,b)$. Now $(a',b')$ is an interval in $K$ containing $d$, so it must also contain a $K$-square. Since $(a',b') \subseteq (a,b)$, this completes the proof.
\end{proof}

\begin{corr}
\label{corr:D-sd-in-Q}
$D$ is square-dense in $Q$ if and only if $D$ is square-dense in itself.
\end{corr}
\begin{proof}
$D$ is dense in $Q$ by \Ax{D}, so the result follows from Corollary \ref{lemma:dense-and-sd}.
\end{proof}

Significantly for our purposes, every square-dense field can be embedded as a \emph{dense subfield} of an algebraic Euclidean ordered extension. We do not know whether the converse is true, nor whether every ordered field is square-dense.

\begin{theorem}
Let $K$ be an ordered field. If $K$ is square-dense, there exists an algebraic extension $L$ of $K$ which is ordered and Euclidean, and in which $K$ is topologically dense.
\end{theorem}
\begin{proof}
By assumption, $K$ is square-dense in its real-closure $R$. Let $L$ be the topological closure of $K$ in $R$. Then $L$ is obviously ordered, and $K$ is dense in $L$. It is easy to see that $K$ must be a field (for example, if $q >0$ in $L$, we can find a convergent net $d_\lambda \to q$ comprising only positive values $d_\lambda \in K$; now $1/d_\lambda \to 1/q$ in $R$, so $1/q \in L$). By the Artin-Schreier theorem, we know that $R$ is algebraic over $K$, and since $K \leq L \leq R$, we see that $L$ is also algebraic over $K$. It only remains to prove that $L$ is Euclidean.

Suppose $q \in L$ is positive. We know that $q$ has a (positive) root in $R$; we'll call it $\sqrt{q}$. We need to show that $\sqrt{q} \in L$. Suppose not. Then, since $L$ is the topological closure of $K$ in $R$, there must be some positive $a,b \in R$ for which $a < \sqrt{q} < b$ and $(a,b) \cap K = \varnothing$. Taking squares, we see that $a^2 < q < b^2$ and $(a^2,b^2)$ contains no $K$-squares. Since $L$ is $K$'s topological closure in $R$, and the $R$-interval $(a^2, b^2)$ contains $q \in L$, there must exist some $d \in K$ with $d \in (a^2, b^2)$. But this means that $(a^2,b^2)$ contains a value from $K$, but no $K$-squares, contrary to hypothesis. Therefore $\sqrt{q} \in L$, and $L$ is Euclidean.
\end{proof}

Since we now know that fields exist with the required properties, we shall henceforth assume 
\begin{quote}
\Ax{D}: $D$ is a square-dense non-Euclidean ordered field, and $Q$ is the topological closure of $D$ in its real-closure.
\end{quote}
We shall refer to a hop \jump{q}{q'} as a $Q$-hop if the coordinates of $q$ and $q'$ all lie in $Q$, and as a $D$-hop if they are restricted to $D$; we define $Q$-paths and $D$-paths analogously.

\subsection{Algebraic hops}
\label{sec:restricting-hops-to-D}

As we have noted, superpositions of $Q$-paths lead inexorably to the continuous motion we expect to see in theories like \SpecRel. But we need to decide \emph{which} hops are permissible. We are arguing that continuous motion in $Q^{N+1}$ is an `illusion' induced by quantum superposition. But if we're happy to assert that \emph{continuous} motion is illusory, what grounds do we have for asserting that finitary motion is \emph{not} illusory? In fact, we argue that \emph{all} motion can be regarded as illusory, and is generated by underspecification of $Q$.

Our starting point is the observation that, because $Q$ is algebraic over $D$, its structure (and hence that of $Q^{N+1}$) is strongly related to the nature of polynomials over $D$. What happens if we only consider swaps that respect these polynomials?

We accordingly say that a $Q$-hop \jump{\Tuple{t',\bar{x}'}} {\Tuple{t',\bar{x}'}} is \emph{algebraic} if the various coordinates (taken in turn) satisfy the same polynomials over $D$, but do not themselves belong to $D$. We consider below what happens if we restrict attention to algebraic hops. We find that the $Q$-viewpoint is essentially unchanged, which suggests that the reason objects are seen to move, and the reason time is seen to flow, is possibly because $Q$ permits `enough' hops which cannot be detected observationally (i.e. at the level of $D$).

Of course, not all hops are algebraic. Nonetheless, it is possible to approximate general hops arbitrarily closely using algebraic ones. We now explain what we mean by this claim.

\subsection{Hop approximations}
\label{sec:hop-approximations}

Suppose $h \equiv \jump{q}{q'}$ is a hop. If we can find algebraic hops carrying $q$'s approximations onto those of $q'$, we will say that those algebraic hops provide an approximation to $h$.

\begin{defn}
We say that a $Q$-hop $\jump{q}{q'}$ can be \emph{approximated} if, given any open $Q$-intervals $(a,b)$ containing $q$ and $(c,d)$ containing $q'$, there exists some
\begin{itemize}
\item
	$\alpha \in Q\setminus D$ in the interval $(a,b)$;
\item
	$\beta \in Q\setminus D$ in the interval $(c,d)$;
\item
	and $\alpha, \beta$ satisfy the same minimal polynomial.
\end{itemize}
\end{defn}

Since the defining feature of hop-based motion in $Q^{N+1}$ is that hops are unrestricted with respect to both their source and their target, our goal in this section is to show that every $Q$-hop can be approximated. Since $D$ is assumed by \Ax{D} to be non-Euclidean, we argue as follows.

\begin{lemma}
\label{lemma:roots-are-dense}
Suppose $K$ is non-Euclidean and dense in some Euclidean extension $L$, and write $\sqrt{K} = \{ q \in L \st q^2 \in K \}$. Then $\sqrt{K} \setminus K$ is also dense in $L$.
\end{lemma}
\begin{proof}
  Let $a < b \in L$, and consider the interval $(a,b) \subseteq L$. Without loss of generality we can assume that $a > 0$. We want to show that $(a,b)$ contains a value from $\sqrt{K} \setminus K$. To begin, we replace the bounds $a$ and $b$ with values from $K$. Since $K$ is dense in $L$, we can find $a', b' \in K$ satisfying $a' \in (a,q)$, $b' \in (q,b)$, so that $a' < q < b'$ and $(a',b') \subseteq (a,b)$. 
  
  Since $L$ is Euclidean and $K$ isn't, there exists some positive $d \in K$ and positive $q \in L \setminus K$ with $q^2 = d$. We will use this value to construct the required value in $(a,b)$.
  
   Since $K$ is dense in $L$ and $0 < a' < b'$, there exists $x \in K$ satisfying $\frac{a'}{b'}q < x < q$. Putting $y = b'x/a'$, we have $y \in K$ and $y = \frac{b'}{a'}x > \frac{b'}{a'}(\frac{a'}{b'}q) = q$, so that $x < q < y$, and hence $a' = \frac{a'}{x}x < \frac{a'}{x}q < \frac{a'}{x}y = \frac{a'}{x}(\frac{b'}{a'}x) = b'$. In other words, if we write $z = \frac{a'}{x}q$, we have $z \in (a',b') \subseteq (a,b)$.
   
   We claim that $z \in \sqrt{K} \setminus K$. To see that $z \in \sqrt{K}$, we note that $z^2 = q^2\left(\frac{a'}{x}\right)^2 = d\left(\frac{a'}{x}\right)^2 \in K$, since $a'$, $d$ and $x \in K$ and $x \neq 0$. On the other hand, we cannot have $z \in K$, since it would then follow that $q = xz/a' \in K$, contrary to hypothesis.
\end{proof}

\begin{corr}
\label{corr:roots-are-dense}
$\sqrt{D} \setminus D$ is dense in $Q$. \QED
\end{corr}

\begin{lemma}
\label{lemma:quadratic-approximation}
Suppose $0 < r < s$ in $Q$, and let $\epsilon \in Q$ satisfy $0 < \epsilon < s-r$. Then there exist $\alpha, \beta \in Q$ satisfying
\begin{itemize}
\item[(i)]
	$\Mod{r - \alpha} < \epsilon$
\item[(ii)]
	$\Mod{s - \beta} < \epsilon$
\item[(iii)]
	$(x - \alpha)(x - \beta) \in D[x]$
\item[(iv)]
	$\alpha, \beta \not\in D$
\end{itemize}
\end{lemma}
\begin{proof}
Since $D$ is dense in $Q$, we can choose $b \in D$ satisfying
\[
	-\frac{r+s}{2} < b < -\frac{r+s}{2} + \frac{\epsilon}{2}
\]
Now $(s+b) - \frac{\epsilon}{2} > (s - \frac{r+s}{2}) -\frac{\epsilon}{2} = \frac{s-r}{2} - \frac{\epsilon}{2} >0$, so the interval $(s+b- \frac{\epsilon}{2}, s+b)$ is entirely positive. By Corollary \ref{corr:roots-are-dense}, we can find some (positive) $z \in \sqrt{D} \setminus D$ in this interval. If we now set $c = (b^2-z^2)/2$, so that $c \in D$, then $b^2 - 2c = z^2$ is positive, $z = \sqrt{b^2 - 2c}$, and
\[
	(s+b) - \frac{\epsilon}{2} < \sqrt{b^2 - 2c} < (s+b)
\]
Let $p(x) \in D[x]$ be the polynomial $p(x) = \frac{1}{2}x^2 + bx + c$, and let $\alpha, \beta$ be its roots in $Q$,
\[
	\alpha = -b - \sqrt{b^2-2c} \qquad \beta = -b + \sqrt{b^2-2c}
\]
Clearly $p = (x-\alpha)(x-\beta)$, whence (iii) is satisfied. Next we note that

\begin{tabular}{ll}
\\
\begin{tabular}{rl}
$r- \alpha$  &$= r + [b + \sqrt{b^2-2c}]$ \\
             &$< (r+b) + (s+b)$ \\
             &$= (r+s) + 2b$ \\
             &$< (r+s) + [\epsilon - (r+s)]$ \\
             &$= \epsilon$
\end{tabular}
&
\begin{tabular}{rl}
$r- \alpha$  &$= r + b + \sqrt{b^2-2c}$ \\
             &$> (r+b) + [(s+b) - \epsilon/2]$ \\
             &$= (r+s) + 2b - \epsilon/2$ \\
             &$> (r+s) - (r+s) - \epsilon/2$ \\
             &$= -\epsilon/2  > -\epsilon$
\end{tabular}
\\
\end{tabular}

\noindent which proves (i), and

\begin{tabular}{ll}
\\
\begin{tabular}{rl}
$s - \beta$  &$= s + [b - \sqrt{b^2-2c}]$ \\
             &$< s+b - (s+b+\epsilon/2)$ \\
             &$= \epsilon/2$ \\
             &$< \epsilon$
\end{tabular}
&
\begin{tabular}{rl}
$s - \beta$  &$= s + [b - \sqrt{b^2-2c}]$ \\
             &$> s+b - (s+b)$ \\
             &$= 0$ \\
             &$> - \epsilon$
\end{tabular}
\\
\end{tabular}

\noindent which proves (ii). Finally, to prove (iv) we note that $z = \sqrt{b^2-2c}= - \alpha - b = \beta - b$. If $\alpha$ or $\beta$ were in $D$, then $z$ would be as well, but $z \in \sqrt{D} \setminus D$.
\end{proof}

Lemma \ref{lemma:quadratic-approximation} tells us that any pair $(r,s)$ of points in $Q$ can be approximated arbitrarily closely by pairs $(\alpha,\beta)$ which are the roots of a minimal quadratic polynomial over $D$. We have therefore proven the result we sought, viz.

\begin{theorem}
\label{thm:every-hop-approximated}
Every $Q$-hop can be approximated. \QED
\end{theorem}

\begin{note} \normalfont{
In the same way that hops can be approximated algebraically, we believe (but have not as yet formally proven) that entire \emph{paths} can be approximated algebraically; that is, given any finite sequence \Tuple{r_j} in $Q$, we can approximate it arbitrarily closely by a sequence \Tuple{\alpha_j} in $Q$, where the $\alpha_j$ are all roots of the same minimal polynomial:
}\end{note}

\begin{quote}
\begin{conj}
\label{conj:path-approx}
Suppose $0 < r_1 < \dots < r_n$ in $Q$, and let $\epsilon \in Q$ satisfy $0 < \epsilon < \min \{ r_j -r_{j-1} \}$. Then there exist $\alpha_1, \dots, \alpha_n \in Q$ satisfying
\begin{itemize}
\item[(i)]
	$\Mod{r_j - \alpha_j} < \epsilon$;
\item[(ii)]
	$(x - \alpha_1)\dots(x - \alpha_n) \in D[x]$;
\item[(iii)]
	$(x - \alpha_1)\dots(x - \alpha_n)$ is irreducible over $D$ \QED
\end{itemize}
\end{conj}
\end{quote}
We discuss evidence for this conjecture in our concluding remarks (section \ref{sec:summary-and-conclusions}).

\subsection{Restricting to algebraic hops}
\label{sec:using-algebraic-hops}

Suppose, then, that we restrict finitary motion to algebraic hops. Do we still find that finitary motion explains the illusion of continuous motion with respect to an apparently `flowing' time? This will remain the case provided we can prove that amplitude calculations can be expressed entirely in terms of algebraic hops, since in that case all physically relevant calculations will yield results identical to those of the bidirectional model, and hence equivalent to those of Feynman's standard formulation. 

\begin{theorem}
\label{thm:main-theorem}
It is possible to define finitary path amplitudes using only algebraic hops.
\end{theorem}

\begin{proof}
Suppose $p \equiv q_0 \to q_1 \to \dots q_n$ is a finitary path in $Q$.

Let $\mathcal{U}$ be the neighbourhood base (in $Q^{N+1}$) at $0$, i.e. $\mathcal{U}$ contains every open $Q^{N+1}$-interval containing $0$. Then because addition is both continuous and bijective, the neighbourhood base at $q_j$ can be obtained in the following obvious way: if $(a,b) \in \mathcal{U}$ we associate it with the interval $(a+q_j, b+q_j)$ that contains $q_j$, and conversely. If we order $\mathcal{U}$ by set-inclusion, it follows that $\mathcal{U}$ is an ordered set which can be used to index the neighbourhoods of each $q_j$ in a consistent way. For each $U \in \mathcal{U}$ we will write $U_j$ for the corresponding interval containing $q_j$. Notice that $\bigcap \mathcal{U} = \{0\}$, whence $\bigcap_{U \in \mathcal{U}} \{ U_j \} = \{ q_j \}$ for each $j$.

Choose any $U \in \mathcal{U}$. By Theorem \ref{thm:every-hop-approximated}, there exist $\alpha_j, \beta_j \in Q \setminus D$ with $\alpha_{j,U} \in U_j$ and $\beta_{j,U} \in U_{j+1}$, and an algebraic hop $g_{j,U}$ taking $\alpha_{j,U}$ to $\beta_{j,U}$.

Define $s_U = \sum_j{ s(\jump{\alpha_{j,U}}{\beta_{j,U}}) }$. By corollary \ref{corr:classical-action-continuous}, we know that $s$ is a continuous function of its endpoints, and because we only ever consider finite sums, we can conclude that
\begin{eqnarray*}
	\lim_{U \in \mathcal{U}} s_U &=& \lim_{U \in \mathcal{U}} \sum_j{ s(\jump{\alpha_{j,U}}{\beta_{j,U}}) } \\
	         &=& \sum_j{\lim_{U \in \mathcal{U}}  s(\jump{\alpha_{j,U}}{\beta_{j,U}}) } \\
	         &=& \sum_j{s(\jump{\lim_{U \in \mathcal{U}}  \alpha_{j,U}}{\lim_{U \in \mathcal{U}}  \beta_{j,U}}) } \\
	         &=& \sum_j{s(\jump{q_j}{q_{j+1}}) } \\
	         &=& s(p)
\end{eqnarray*}

Consequently, the path action $s(p)$ can be replaced in the bidirectional formulation by the equivalent expression $\lim_{U \in \mathcal{U}} \sum_j{ s(\jump{\alpha_{j,U}}{\beta_{j,U}}) }$. Since all of the hops $\jump{\alpha_{j,U}}{\beta_{j,U}}$ used to construct this expression are algebraic, and the amplitude of the path $p$ is defined by $B_n e^{-is(p)/\hbar}$, it follows (as claimed) that this amplitude also can be defined without reference to non-algebraic hops.
\end{proof}

\section{Summary and Conclusions}
\label{sec:summary-and-conclusions}

\paragraph{Why objects appear to move:}
It follows immediately from Theorem \ref{thm:main-theorem} that the equations of finitary motion can all be re-expressed in terms of algebraic hops, whence the existence of algebraic hops is itself enough to generate the `illusion' that particles move along continuous paths in $Q^{N+1}$. In particular, since it tells us that $m$ sees herself moving continuously along her time axis, it also tells us why time appears to flow: continuous motion is an artefact of the model itself, and comes about because of our mistaken use of $Q$ to coordinatize spacetime instead of $D$. As we have seen, there necessarily exist algebraic hops of $Q$ which leave $D$ fixed pointwise (i.e. they have no `real' effect), and these are sufficient to approximate all other hops. Consequently, `reality' is required to be invariant under the action of these hops (and the illusions they generate) --- in effect, `physical laws of motion' reflect the properties of minimal polynomials over $D$.

Are we justified, therefore, in arguing that motion appears to be continuous --- and time appears to flow --- \emph{because} algebraic hops are physically realisable? The answer, as always, depends on our terms of reference. Certainly, if we accept that theories like \SpecRel and its extensions provide a valid description of physical reality (in whatever context), then it would appear that continuous motion is a necessary consequence of the existence of algebraic hops. But the existence theorem for these hops has several assumptions built-in, and these need to be considered carefully. We have assumed, in particular, that observations are made in some field $D$ which is square-dense but not Euclidean. While this appears to be valid, in the sense that `real-world' measurements appear to be made in a \QQ-like field, this is not \emph{proof} that a larger field will never one day be needed.

Our construction also tells us how to identify a suitable field $Q$ for use in \SpecRel. We should take the topological closure of $D$ in its real-closure. But we have no explanation for the \emph{psychology} of this choice. In what sense can the procedure `take the topological closure of the field in which practical measurements are made' be considered a natural intuition? Indeed, this rule would tell us to use the field of algebraic reals to reason about space and time, yet the normal choice is the field \RR of \emph{all} reals.

\paragraph{A potential application (semi-traversable wormholes):}
As Novikov has explained, certain types of wormhole can be expected to collapse in such a way as to become semi-traversable: even though a signal has successfully managed to cross the wormhole before it collapsed, this signal is trapped in the throat of the wormhole and cannot escape into the target universe. One way to access the signal is to send exotic matter into the wormhole, so as to reduce its gravitational radius; but according to our model it should also be possible for the signal to emerge spontaneously, because we do not require photons to follow continuous paths: any given photon can jump from \emph{any location to any other}, even if that location is beyond an event horizon or located in another universe. Since the amplitude of such a hop is given by $e^{-is/\hbar}$, the only way to prevent it happening is to set $s = +\infty$, and it is unclear whether this would ever be meaningful. In particular, we would expect a finite action to be associated with the possibility that the particle hops from one side of the wormhole to the other `the long way round'; then, since the particle would seem to have appeared at the target end of the wormhole faster than it could possibly have travelled via the `standard' route, an observer within the system would naturally conclude that it had indeed passed through the collapsed wormhole.

However, even this language is misleading. According to our model, particles hop directly from one location to another, without passing through intervening points. Statements of the form ``the signal travelled through the wormhole'' are, therefore, essentially meaningless. Rather, we would say that the existence of the wormhole introduces additional finitary paths between any two points (since paths can now include hops to and from points \emph{within} the wormhole).

\paragraph{Further work:}
The work presented here is at a very early stage. We have shown that individual hops can be approximated using `quadratic hops', but the question naturally arises whether other hops are also possible. In particular, is Conjecture \ref{conj:path-approx} valid? If we have a path $q_0 \to \dots \to q_n$, is it possible to find points $\alpha_j$ arbitrarily close to the $q_j$, which all happen to be roots of the same minimal polynomial? Our proof of Lemma \ref{lemma:quadratic-approximation} involved the direct construction of an irreducible quadratic, and relied on our ability to construct the radical $\sqrt{b^2-2c}$ associated with the polynomial $\frac{1}{2}x^2 + bx +c$. In order to construct suitable roots for polynomials of higher degree, it is likely that we will need to consider radical-based methods for solving polynomials, and it is well-known that these are not always available. Nonetheless, it may be possible to use specific polynomials of sufficiently high degree, and it may be possible to adapt standard solvability results to identify which path lengths always permit approximation.

It is unclear, however, what the physical significance of such a solution would be. We have already seen that quadratic solutions are sufficient to replicate the whole of the bidirectional model's physical manifestations; having other solutions available may perhaps tell us something about the structure and topology of `apparent' spacetime.

\section*{Acknowledgement}
This research was funded in part by the EPSRC. My especial thanks go to the members of the Budapest Relativity Group (especially \Istvan, Hajnal, Judit and Gergely), who very kindly set aside several days in early 2009 to explain their logical approach to relativity theory to me.

\bibliography{budapest}
\end{document}